# Rethinking Factor Loading Thresholds:
# A Case for a Strict λ ≥ .70 Rule

**Authors;**
M.Murat YAŞLIOĞLU, Istanbul University, School of Business (Corresponding Author)
Duygu Toplu YAŞLIOĞLU, Istanbul University, School of Business

## Abstract

This paper challenges the prevailing practice of accepting standardized factor loadings as low as .50 in confirmatory factor analysis. Drawing on the logic of Average Variance Extracted (AVE) and communality, the author argues for a stricter item level threshold: only indicators with loadings of $\lambda \geq .70$ (implying $\lambda^2 \geq .50$) should be retained in final measurement models. The rationale is that indicators with $\lambda < .70$ contain more error than explained variance, undermining both construct validity and the stability of factor solutions. The paper reviews theoretical foundations, simulation evidence, and implications for structural equation modeling, showing that weak loadings degrade measurement quality, factor score determinacy, and model fit. Adopting a minimum $\lambda \geq .70$ rule aligns item level standards with established construct level criteria and enhances the rigor and interpretability of latent variable models.

**Keywords**: factor loadings, loading tresholds, factor analysis, confirmatory factor analysis

## Introduction

Factor loadings are central to the interpretation of latent variable models in both exploratory factor analysis (EFA) and confirmatory factor analysis (CFA). A standardized factor loading (λ) represents the strength of the relationship between a latent factor and an observed indicator; its square, $\lambda^2$, is the proportion of variance in the indicator explained by the factor. The remaining variance, $(1 - \lambda^2)$, is unique or error variance.

Prevailing guidelines typically recommend that standardized loadings in CFA should be at least .50 and ideally .70 or higher (Hair et al., 2018). Convergent validity is usually evaluated at the construct level using the Average Variance Extracted (AVE), with AVE ≥ .50 taken to mean that a latent construct explains at least half of the variance in its indicators (Fornell & Larcker, 1981).

This paper proposes a stricter standard: any reflective indicator with $|\lambda|$ below .70 should not be accepted in final CFA solutions. A loading below this threshold means that most of the indicator's variance is due to error. This position applies the logic of AVE and communality consistently at the item level.

## Conventional Thresholds and the AVE Logic

In mainstream SEM practice, the measurement quality of a construct is commonly assessed via three related criteria: factor loadings, composite reliability, and AVE. Hair et al. (2018) and numerous applied studies repeat the rule that standardized loadings should be ≥ .50, and ideally ≥ .70, and that AVE should be ≥ .50 to support convergent validity. Fornell and Larcker's (1981) original formulation is explicit: when AVE < .50, more error remains in the items than variance explained by the latent construct (Cheung et.al. 2024).

AVE is defined as:

$AVE = (1/k) \Sigma \lambda_i^2$

where k is the number of indicators. AVE ≥ .50 therefore implies that, on average, at least half of each indicator's variance is accounted for by the latent variable. AVE < .50, by this same definition, implies that the average indicator is dominated by error.

It is important to make explicit that this logic also applies to each indicator. For item i, $\lambda_i^2$ shows the proportion of variance explained by the factor, while $(1 - \lambda_i^2)$ represents error or unique variance. If $\lambda_i^2$ is less than .50, then the item contains more error than signal in relation to the construct.

## A Consistent Item Level Criterion: $\lambda^2 \geq .50$

Consider the standard reflective measurement model for a standardized item $x_i$:

$x_i = \lambda_i \xi + \varepsilon_i$, with $Var(\xi) = Var(x_i) = 1$.

Then:

- Communality for the item: $h_i^2 = \lambda_i^2$
- Unique (error) variance: $\theta_i = 1 - \lambda_i^2$

Numerically:

- $\lambda = .50 \rightarrow h^2 = .25$, 75% error
- $\lambda = .60 \rightarrow h^2 = .36$, 64% error
- $\lambda = .70 \rightarrow h^2 = .49$, 51% error
- $\lambda = .80 \rightarrow h^2 = .64$, 36% error

The AVE criterion requires AVE ≥ .50, which means that, on average, the construct should explain at least half of the variance in its indicators. If we follow this logic at the item level, it does not make sense to accept indicators with $\lambda^2$ below .50. These indicators have more variance due to error than to the latent factor.

From this reasoning, the following rule can be stated:

$\lambda^2 \geq .50 \Rightarrow |\lambda| \geq \sqrt{.50} \approx .707$.

To summarize, if AVE values below .50 are not accepted at the construct level because they show more error than explained variance, then indicators with $\lambda^2$ below .50 should also not be accepted for the same reason. Allowing items with $\lambda$ below .70 while requiring AVE ≥ .50 overlooks weaknesses at the item level and can conceal them.

## Communalities, and Factor Stability

The same argument appears in EFA through the concept of communality. For item i, communality is the sum of squared loadings across all factors. In a simple structure solution where an item loads meaningfully on only one factor, $h_i^2 = \lambda_i^2$.

Simulation work by Guadagnoli and Velicer (1988) and subsequent studies shows that factor solutions are unstable when communalities are low, and that strong loadings and higher communalities dramatically improve factor stability, especially in moderate samples. Recommendations frequently suggest that communalities below about .40 are problematic and that factors are most reliable when there are several items with loadings $\geq .60$.

If we use a communality standard similar to the AVE standard, such as $h^2 \geq .50$ for items to be adequately represented by a factor, we again reach the criterion $\lambda^2 \geq .50$, or $\lambda \geq .70$. Items with $\lambda$ below .70 have $h^2$ below .50, so they also contain more error than explained variance in CFA.

For these reasons, applying a stricter $\lambda \geq .70$ rule in final CFA solutions is a logical extension of the AVE argument to the item level, using the concept of communality.

## Structural and Model Fit Consequences of Weak Loadings

It is sometimes argued that CFA adjusts for measurement error, so low loadings are less concerning. However, weak indicators still reduce the effectiveness of latent variable models.

First, low loadings reduce factor score determinacy: indicators with weak loadings carry little information about the factor, which leads to noisier latent variable estimates and less precise structural path coefficients.

Second, widely used fit index cutoffs (e.g., CFI, TLI, RMSEA) were originally evaluated and calibrated under conditions where factor loadings were relatively high (often around .70–.80) in simulation studies such as Hu and Bentler (1999). As more recent work on fit indices emphasizes, the performance of fit indices depends on measurement quality. Applying standard fit cutoffs to models dominated by weak loadings quietly stretches those guidelines beyond their empirical justification.

Third, the EFA literature shows that when communalities are low, factor recovery is unstable and highly sensitive to sample fluctuations (Guadagnoli & Velicer, 1988). Even with latent variable SEM, a measurement model built on mostly weak indicators yields fragile structural inferences: the latent variable is simply too noisy to serve as a robust explanatory construct.

Taken together, these points indicate that accepting indicators with $\lambda$ well below .70 is a significant concern. Including such items reduces the quality of measurement and structural estimates and may create a misleading impression of adequacy based on global fit indices or AVE averages.

## A Normative Measurement Rule

The proposed position can be summarized in three steps:

1. Axiomatic acceptance

The field broadly accepts that AVE $\geq .50$ is a necessary condition for convergent validity: a construct must, on average, explain at least half the variance of its indicators (Fornell & Larcker, 1981).

2. Consistency requirement

For each individual indicator, $\lambda^2$ plays the same conceptual role as AVE at the construct level. If $\lambda^2 < .50$, then the item is, by definition, dominated by error relative to that construct.

3. Normative rule

Therefore, in final CFA solutions that are used for interpretation, indicators with |λ| below .70 should not be accepted. They do not meet the same 'more variance explained than error' principle that AVE ≥ .50 represents at the construct level.

It is important to note that in early exploratory analyses or when practical limitations exist, researchers may need to accept lower loadings. Still, as a measurement standard, the λ ≥ .70 rule offers a clear and consistent baseline that aligns item level practice with the theoretical principles of AVE and communality.

## Conclusion

Current guidelines already state that λ around .70 is ideal and AVE ≥ .50 is needed for convergent validity. The argument here extends these principles: if we do not accept constructs whose indicators are, on average, more error than signal, we should also not accept individual indicators that are more error than signal.

In conclusion, a factor loading below .70 is not only less than ideal but also inconsistent with established measurement theory. Treating λ ≥ .70 as a minimum requirement, rather than only a target, can help make factor analysis more transparent and rigorous.